\documentclass{elsart}

\usepackage{amsmath,amsfonts}

\newcommand{\thmref}[1]{Theorem~\ref{#1}}
\newcommand{\lemref}[1]{Lemma~\ref{#1}}

\let\abs=\envert

\begin{document}
\begin{frontmatter}
\title{Proximity Inversion Functions on the Non-Negative Integers}
\author{Brendan Lucier}
\address{School of Computer Science, University of Waterloo, Waterloo, Ontario, Canada, N2L 3G1}

\begin{abstract}
We consider functions mapping non-negative integers to non-negative real numbers such that $a$ and $a+n$ are mapped to values at least $\frac{1}{n}$ apart. In this paper we use a novel method to construct such a function. We conjecture that the supremum of the generated function is optimal and pose some unsolved problems.
\end{abstract}

\end{frontmatter}

\section{Introduction}

In the Constraint Satisfaction Problem, one is given a set of variables and must find an assignment of values that respects certain constraints. For a general survey of constraint satisfaction, see \cite{jeav:csp}. A subset of constaint satisfaction problems is the binary-constraints problem (BCP), wherein each constraint affects only two variables. This problem has theoretical significance but can also be applied to many practical problems, such as frequency assignment \cite{dunkin:freq}.

We consider a generalization of the BCP where the set of variables is taken to be an arbitrary metric space. We shall call this problem the Metric Space BCP (MSBCP). This generalization was first formulated in \cite{flaass:freq}. If $M$ is the set of variables in an instace of the MSBCP, then a solution can be expressed as a function $f\colon M \to \Rset^{\geq 0}$ that satisfies certain constraints. We call such a function an inverse proximity function. We specify this definition more formally in Section \ref{sec:proxinv}.

There are infinitely many solutions to any MSBCP, so we introduce the notion of an optimal solution. Given all solutions for an instance of the MSBCP, an optimal solution $f^*$ is one that minimizes $\sup_{a\geq0}f(a)$. Heuristically, we consider such a function optimal because it satisfies the constraints in as little space as possible.

In this paper we shall examine a particular instance of the MSBCP. Given a function $f\colon \Nset \to \Rset^{\geq 0}$, we require that $\abs{f(a)-f(a+n)} \geq \frac{1}{n}$ for all $a \geq 0$ and $n \geq 1$. We are interested in minimizing the supremum of such a function. 

Fon-Der-Flaass \cite{flaass:freq} conjectured that the supremum of a function satisfying the above constraints could be no less than $1 + \phi$, where $\phi = \frac{1+\sqrt{5}}{2}$ is the golden ratio. In this paper we shall construct a particular function $f$ that satisfies the required criteria such that
\[ \sup_{a \geq 0}f(a) = 1 + \sum^\infty_{\,n \geq 1}\frac{1}{F_{2n}} \]
where $F_n$ denotes the $n^{th}$ Fibonacci number. We have no closed-form expression for this limit, but its value is known to be approximately $2.5353\dotsc < 1 + \phi$.

We believe that our solution is optimal, but this has yet to be proved. More importantly, we believe that our method for constructing the solution is generalizable to other instances of the MSBCP. It is our hope that this will lead to a general method for constructing optimal (or near-optimal) solutions.

\section{Definitions and Previous Work}

\subsection {Fibonacci Numbers}

We shall take $\Nset = \{0, 1, 2, 3, \dotsc \}$ to be the {\em natural numbers\/}. Recall that the {\em Fibonacci sequence\/} is a sequence of natural numbers defined by $F_0 = 0$, $F_1 = 1$, and $F_n = F_{n-1} + F_{n-2}$ for all $n > 1$. The entries of the Fibonacci sequence are referred to as the {\em Fibonacci numbers\/}.
Also recall Catalan's Identity \cite{morgado:catalan}:
\begin{equation} \label{catalan}
F_n^2 - F_{n+r}F_{n-r} = (-1)^{n+r}F_r^2 \text{ for all } 0 \leq r \leq n.
\end{equation}

\subsection {Strings}

An {\em alphabet\/} is a non-empty (possibly infinite) set of characters. Given an alphabet $\Sigma$, a {\em word\/} or {\em string\/} over $\Sigma$ is a (finite or infinite) sequence of characters of $\Sigma$. We write $\Sigma^n$ to mean the set of all words over $\Sigma$ of length $n$. Denote by $\Sigma^*$ the set of all finite words over $\Sigma$. 

Given a word $x$, we shall write $x[i]$ to denote the $i$th character of $x$. We shall take indexing to start at $1$, so $x = x[1]x[2]x[3]\dotsm$. We shall write $x^*$ to mean zero or more occurances of $x$, $x^n$ to mean exactly $n$ occurances of $x$, and, for any other word $y$, $xy$ to represent the word consisting of $x$ followed by $y$.

\subsection {Numeration Systems}

A {\em numeration system\/} is specified by a set $S = \{u_1, u_2, u_3, \dotsc \}$ of strictly increasing natural numbers with $u_1 = 1$. The following theorems are due to Fraenkel \cite{fraen:numsys}.

\begin{thm}\label{fraen:thm1}
Any nonnegative integer $N$ has precisely one representation in the system $S = \{u_1,u_2,\dotsc\}$ of the form $N = \displaystyle\sum_{i=1}^nd_iu_i$, where the $d_i$ are nonnegative integers satisfying
\begin{equation} \label{fraen:eq1}
d_iu_i + d_{i-1}u_{i-1} + \dotsm + d_1u_1 < u_{i+1}
\end{equation}
for all $i > 0$.
\end{thm}

\begin{thm} \label{fraen:thm2}
For $m \geq 1$, let $b_1, b_2, \dotsc$ be integers satisfying 
\[ 1 \leq b_m \leq \dotsm \leq b_2 \leq b_1. \]
Let $u_{-m+1}, u_{-m+2}, \dotsc, u_{-1}$ be fixed nonnegative integers, and let 
\[ u_0 = 1, u_n = b_1u_{n-1} + b_2u_{n-2} + \dotsm + b_mu_{n-m} \]
for all $n \geq 1$. Then any nonnegative integer $N$ has precisely one representation in $S = \{u_i\}$ of the form $N = \displaystyle\sum_{i=0}^nd_iu_i$ if the digits $d_i$ are nonnegative integers satisfying the following (two-fold) condition:
\begin{enumerate}
\renewcommand{\labelenumi}{(\roman{enumi})}
\item Let $k \geq m-1$. For any $j$ satisfying $0 \leq j \leq m-2$, if 
\begin{equation}\label{fraen:eq}
(d_k,d_{k-1},\dotsc,d_{k-j+1}) = (b_1,b_2,\dotsm,b_j),
\end{equation}
then $d_{k-j} \leq b_{j+1}$; and if \eqref{fraen:eq} holds with $j=m-1$ then $d_{k-m+1} < b_m$.
\item Let $0 \leq k < m-1$. If \eqref{fraen:eq} holds for any $j$ satisfying $0 \leq j \leq k-1$, then $d_{k-j} \leq b_{j+1}$; and if \eqref{fraen:eq} holds with $j = k$, then $d_0 < \displaystyle\sum_{i=k+1}^mb_iu_{k+1-i}$.
\end{enumerate}
\end{thm}

Fraenkel \cite{fraen:numsys} also shows that the representation in \thmref{fraen:thm2} satisfies \eqref{fraen:eq1}.

Over a given numeration system $S$, we can express the unique  representation of an integer $N$ as the word $d_1d_2\dotsm d_n$ over the alphabet $\Nset$. In a slight abuse of notation, we shall also refer to any word of the form $d_1d_2\dotsm d_n0^*$ as {\em a representation of $N$\/}. In general, given any word $x \in \Nset^*$, we say that $x$ is a {\em valid represenation\/} if there exists some $N$ for which $x$ is a representation for $N$. That is, $x$ is a valid representation if and only if the digits of $x$ satisfy the conditions of \thmref{fraen:thm2} (since any trailing zeros will not violate the constraints of \thmref{fraen:thm2}).

\subsection{Proximity Inversion Functions}
\label{sec:proxinv}

A {\em constraint function\/} is a non-increasing function $c \colon \Rset^{>0} \to \Rset^{\geq 0}$. Recall that a {\em metric space} consists of a set $M$ and a distance function $d \colon M \to \Rset^{\geq 0}$ such that
\begin{enumerate}
\item $d(a,b) \geq 0$
\item $d(a,b) = 0 iff a = b$
\item $d(a,b) = d(b,a)$
\item $d(a,b) = d(a,c) + d(c,b)$
\end{enumerate}
for all $a,b,c \in M$. 

The {\em Metric Space Binary Constraints Problem\/} (MSBCP) is as follows: given an arbitrary metric space $(M,d)$ and constraint function $c$, find a function $f\colon M \to \Rset^{\geq 0}$ satisfying $\abs{f(a)-f(b)} \geq c(d(a,b))$ for all distinct $a,b \in M$. We call the function $f$ a {\em proximity inversion function on $(M,d)$ over $c$\/}.

In this paper we limit ourselves to the metric space of non-negative integers $\Nset$, under the metric
\[ d_\Nset(a,b) = \abs{a-b}. \]
Note that the range of $d_\Nset$ is $\Nset$, so a constraint function for $d_\Nset$ need only be defined over $\Nset^{>0}$. In particular, we wish to find a proximity inversion function on the non-negative integers over the constraint function $c(n) = \frac{1}{n}$. 

\section{Construction}

Our function $f$ will be based upon a Fibonacci numeration system. Let $u_i = F_{2i}$ for all $i \geq 1$. Note that $u_1 = 1$ and $u_i < u_{i+1}$ for all $i \geq 1$. We can therefore consider the numeration system $S = \{ u_i \}_{i=1}^\infty$.

\begin{thm}\label{fibrep}
Any $N \geq 0$ has a unique representation of the form $N = \displaystyle\sum_{i=1}^\infty d_iu_i$, where 
\begin{enumerate}
\renewcommand{\labelenumi}{(\roman{enumi})}
\item $0 \leq d_i \leq 2$ for all $i \geq 1$
\item $d_i = 0$ for all but finitely many values of $i$
\item if $i < j$ and $d_i = 2 = d_j$ then there exists some $l$, $i < l < j$, such that $d_l = 0$.
\end{enumerate}
\end{thm}
\begin{pf}
If $N = 0$ then take $d_i = 0$ for all $i$. This is a unique representation that satisfies the required properties.

Consider $N > 0$. Let $n$ be defined as in \thmref{fraen:thm1}. Property $(i)$ follows from \eqref{fraen:eq1} plus the fact that $3F_{2k} > F_{2k+2}$ for all $k > 1$. Property $(ii)$ follows from \thmref{fraen:thm1}, since we must have $d_k = 0$ for all $k > n$.

For property $(iii)$ note that, given any $n > 0$, $F_{2n} = 2F_{2n-2} + \displaystyle\sum_{i=1}^{n-2}F_{2i} + 1$. So take $u_0 = 1$, $u_i = 0$ for all $i < 0$, $b_1 = 2$, and $b_i = 1$ for all $i > 1$.

If we take $m = n+1$, we get
\[ u_1 = 1,\, u_n = \displaystyle\sum_{i=1}^{m}b_iu_{n-i} \]
for all $n > 1$. \thmref{fraen:thm2} then gives us that $N$ has a unique representation of the form $\displaystyle\sum_{i=1}^nd_iu_i$, where for any $i$,$j$ satisfying $0 < i \leq j \leq n$, if
\[ (d_j, d_{j-1}, \dotsc, d_i) = (2, 1, 1, \dotsc, 1), \]
then $d_{i-1} < 2$. This is equivalent to condition $(iii)$. Note that the numeration system from \thmref{fraen:thm2} corresponds to $S$ for entries less than $N$, but may be different for entries greater than $N$. However, the representation for $N$ will be the same in both systems, so we need not be concerned.
\qed
\end{pf}

Given $a \in \Nset$, let $d^a_i$ denote $d_i$ in the representation of $a$ from \thmref{fibrep}. Also from \thmref{fibrep} there must exist a minimal $l \geq 0$ such that $d^a_j = 0$ for all $j > l$. We shall call this minimal $l$ the {\em length of $a$\/} and write it as $L(a)$.

\thmref{fibrep} also implies that $x \in \{0,1,2\}^*$ is a valid representation with respect to $S$ iff $x$ does not contain a subword of the form $21^*2$. From now on, we shall write $a \equiv x$ to mean that $x$ is a representation of $a$ with respect to $S$. Note that $0 \equiv \epsilon$.

We are now ready to construct our function.
Define $f\colon \Nset \to \Rset^{\ge 0}$ by
\begin{equation}\label{construct}
f(a) = \displaystyle\sum_{i=1}^\infty \frac{d^a_i}{u_i}.
\end{equation}
That is, $f$ applies the digits of an integer's represention in $S$ to the reciprocals of the elements of $S$.

\section{Preliminary Lemmas}

\subsection{Fibonacci Inequalities}
\label{sec:fib}

Before proving that $f$ satisfies the properties we required, we shall need a series of technical lemmas regarding Fibonacci numbers. These lemmas give us properties of our numeration system $\{u_i\}$ that will be useful later.

\begin{lem}\label{catfrac}
Let $r$ and $k$ be integers such that $0 \leq r < k$. Then $\frac{F_{k+r}}{F_{k}^2-1} \leq \frac{1}{F_{k-r}}$ iff $k+r$ is even, with equality occuring when $r \in \{1,2\}$.
\end{lem}
\begin{pf}
Suppose $k+r$ is even. By \eqref{catalan}, $F_{k+r}F_{k-r} = F_{k}^2-F_{r}^2 \leq F_{k}^2-1$ with equality occuring iff $r = 1$ or $r = 2$.\\
Suppose instead that $k+r$ is odd. By \eqref{catalan}, $F_{k+r}F_{k-r} = F_{k}^2+F_{r}^2 > F_{k}^2-1$.
\qed
\end{pf}

\begin{lem}\label{fibsumsimple}
$\displaystyle\sum_{i=k}^nF_{2i} = F_{2n+1} - F_{2k-1}$ for all $0 < k \leq n$.
\end{lem}
\begin{pf}
By induction on $n-k$. If $n=k$, then $F_{2k} = F_{2k+1} - F_{2k-1}$ as required.\\
If $n-k=t>0$ and we assume the result is true whenever $n-k<t$, then
\begin{equation}
\displaystyle\sum_{i=k}^nF_{2i} = F_{2k} + \displaystyle\sum_{i=k+1}^nF_{2i} = F_{2k} + (F_{2n+1} - F_{2k+1}) = F_{2n+1} - F_{2k-1}
\end{equation}
by induction.
\qed
\end{pf}

\begin{cor}\label{fibsumsimplecor1}
If $n > k+1$ then $u_k + u_n - \biggl(\displaystyle\sum_{i=k+1}^{n-1}u_i\biggr) - u_{k+1} - u_{n-1} = 0$.
\end{cor}

\begin{cor}\label{fibsumsimplecor2}
If $n > k+1$ then $2u_{k+1} + \biggl(\displaystyle\sum_{i=k+2}^nu_i\biggr) - u_k = F_{2n+1}$.
\end{cor}

\begin{cor}\label{fibsumsimplecor3}
If $n > k$ then $u_n - u_{n-1} - \displaystyle\sum_{i=k}^{n-1}u_i = F_{2k+1}$.
\end{cor}

\begin{lem}\label{fibsum}
Suppose $1 < k \leq n$. Then
\[ \sum_{i=k}^n\frac{1}{F_{2i}} < \frac{1}{F_{2k-2}} + \frac{1}{F_{2n+2}} - \frac{1}{F_{2k}} - \frac{1}{F_{2n}}. \]
\end{lem}
\begin{pf}
By induction on $n-k$. Suppose first that $n = k$. Note that
\[
\frac{1}{F_{2k}} + \frac{2}{F_{2k}} - \frac{1}{F_{2k+2}}
= \frac{F_{2k+4}}{F_{2k+1}^2-1}
< \frac{1}{F_{2k-2}}
\]
by \lemref{catfrac}. This proves the base case.

Suppose now $n-k=l>0$ and the result is true whenever $n-k<l$. Using induction, we have
\begin{equation*}
\begin{split}
& \biggl(\displaystyle\sum_{i=k}^n\frac{1}{F_{2i}}\biggr) + \frac{1}{F_{2k}} + \frac{1}{F_{2n}} - \frac{1}{F_{2n+2}}\\
= & \biggl[ \biggl( \displaystyle\sum_{i=k+1}^{n}\frac{1}{F_{2i}} \biggr) + \frac{1}{F_{2k+2}} + \frac{1}{F_{2n}} - \frac{1}{F_{2n+2}} \biggr] - \frac{1}{F_{2k+2}} + \frac{2}{F_{2k}}\\
< & \frac{1}{F_{2k}} - \frac{1}{F_{2k+2}} + \frac{2}{F_{2k}}\\
< & \frac{1}{F_{2k-2}},
\end{split}
\end{equation*}
as in the base case.
\qed
\end{pf}

\begin{cor}\label{fibsumcor1}
If $n > k+1$ then $\biggl(\displaystyle\sum_{i=k+1}^{n-1}\frac{1}{u_i}\biggr) + \frac{1}{u_{k+1}} + \frac{1}{u_{n-1}} \leq \frac{1}{u_n} + \frac{1}{u_k}$.
\end{cor}

\begin{cor}\label{fibsumcor2}
If $n > k+1$ then $\biggl(\displaystyle\sum_{i=k+1}^n\frac{1}{u_i}\biggr) + \frac{1}{u_{k+1}} + \frac{1}{F_{2n+1}} < \frac{1}{u_k}$.
\end{cor}
\begin{pf}
Applying \lemref{fibsum} and \lemref{catfrac}, we get
\begin{equation*}
\begin{split}
& \biggl(\displaystyle\sum_{i=k+1}^n\frac{1}{u_i}\biggr) + \frac{1}{u_{k+1}} + \frac{1}{F_{2n+1}}\\
< & \frac{1}{u_k} + \frac{1}{u_{n+1}} - \frac{1}{u_n} + \frac{1}{F_{2n+1}}\\
\leq & \frac{1}{u_k} - \frac{F_{2n+1}}{F_{2n+1}^2-1} + \frac{1}{F_{2n+1}}\\
< & \frac{1}{u_k}
\end{split}
\end{equation*}
as required.
\end{pf}

\begin{lem}\label{fibsum2}
If $n > k$ then
\[
\frac{1}{F_{2n}} + \frac{1}{F_{2k-1}} \leq \biggl(\displaystyle\sum_{i=k}^{n-2}\frac{1}{F_{2i}}\biggr) + \frac{2}{F_{2n-2}}.
\]
\end{lem}
\begin{pf}
By induction on $n-k$. If $n = k + 1$, the claim becomes
\[ \frac{1}{F_{2k+2}} + \frac{1}{F_{2k-1}} \leq \frac{2}{F_{2k}}, \]
but
\[ \frac{2}{F_{2k}} - \frac{1}{F_{2k+2}} = \frac{F_{2k+3}}{F_{2k+1}^2-1} \geq \frac{1}{F_{2k-1}}
\]
by \lemref{catfrac} as required.

If we suppose $n-k > 1$, then by induction we get
\begin{equation}
\begin{split}
& \biggl(\displaystyle\sum_{i=k}^{n-2}\frac{1}{F_{2i}}\biggr) + \frac{2}{F_{2n-2}}\\
= & \biggl(\displaystyle\sum_{i=k+1}^{n-2}\frac{1}{F_{2i}}\biggr) + \frac{2}{F_{2n-2}} + \frac{1}{F_{2k}}\\
\geq & \frac{1}{F_{2n}} + \frac{1}{F_{2k+1}} + \frac{1}{F_{2k}}\\
\geq & \frac{1}{F_{2n}} + \frac{1}{F_{2k-1}}
\end{split}
\end{equation}
by \lemref{catfrac} as required.
\qed
\end{pf}

\subsection{Relative Ordering}

Given two integers represented in decimal notation with equal numbers of digits, one can easily determine which is greater by scanning the digits of the numbers from left to right. This notion extends to general numeration systems as well, as given by the following proposition.

\begin{prop}\label{maxdig}
Take $a,b \geq 0$, $a \neq b$. Since $d^a_i = 0 = d^b_i$ for all $i > \max\{L(a), L(b)\}$, we can find a maximal $l$ such that $d^a_l \neq d^b_l$. Then $a < b \iff d^a_l < d^b_l$. 
\end{prop}
\begin{pf}
This follows directly from \eqref{fraen:eq1}.
\end{pf}

We wish to develop a similar test for the relative ordering of $f(a)$ and $f(b)$. In particular, we shall prove the following theorem.

\begin{thm}\label{invmax}
Given $a,b \geq 0$, $a \neq b$, let $l$ be the minimal value such that $d^a_l \neq d^b_l$. Then $f(a) < f(b) \iff d^a_l < d^b_l$.
\end{thm}
\begin{pf}
Suppose $d^a_l < d^b_l$. Then $d^b_l > 0$, so $L(b) \geq l$.\\
We proceed by induction on $L(a) - l$. Let
\[ c = \sum_{i=1}^ld^b_iu_i. \]
Then $f(c) \leq f(b)$, so it is sufficient to show that $f(a) < f(c)$.\\
If $L(a) \leq l$ then we have that $0 = d^a_i \leq d^c_i$ for all $i > l$. Note also that $d^a_l < d^c_l$ and $d^a_i = d^c_i$ for all $i < l$. Therefore
\[ f(a) = \sum_{i=1}^\infty \frac{d^a_i}{u_i} < \sum_{i=1}^\infty \frac{d^c_i}{u_i} = f(c). \]
Now suppose $L(a) = l + k$, $k \geq 1$. Choose $x \in \{0,1,2\}^{L(a)-l}, y \in \{0,1,2\}^l$ such that $a \equiv yx$. Let
\[ z = y21^{k-1} \in \{0,1,2\}^{L(a)} \]
and suppose first that $z$ is a valid representation. Take $p$ such that $p \equiv z$. We then have 
\begin{equation}\label{eqarg1}
\begin{split}
f(p) & = \sum_{i=1}^{L(a)}\frac{d^p_i}{u_i}\\
& = \biggl(\sum_{i=1}^{l-1}\frac{d^a_i}{u_i}\biggr) + \frac{d^a_l}{u_l} + \biggl(\sum_{i=l+1}^{L(a)}\frac{1}{u_i}\biggr) + \frac{1}{u_{l+1}}\\
& \leq \biggl(\sum_{i=1}^{l-1}\frac{d^b_i}{u_i}\biggr) + \frac{d^b_l-1}{u_l} + (\frac{1}{u_l} + \frac{1}{u_{L(a)+1}} - \frac{1}{u_{L(a)}} - \frac{1}{u_{l+1}}) + \frac{1}{u_{l+1}}\\
& = \biggl(\sum_{i=1}^{l}\frac{d^c_i}{u_i}\biggr) + \frac{1}{u_{L(a)+1}} - \frac{1}{u_{L(a)}}\\ 
& < \sum_{i=1}^{l}\frac{d^c_i}{u_i}\\
& = f(c).
\end{split}
\end{equation}
Now if $a = p$ then $f(a) = f(p) < f(c)$ as required.\\
If $a \neq p$ then $x \neq z$, so there must be a minimal $j > l$ such that $x[j] \neq z[j]$. But if $x[j] > z[j]$ then $x$ must have a prefix of the form $y21^*2$ which contradicts $x$'s validity. Thus $x[j] < z[j]$. But $L(a) - j < L(a) - l$, so by induction we find that $f(a) < f(p) < f(c)$ as required.

Suppose that $z$ is not a valid derivation. Then $y1^t21^{k-t-1}$ is not valid for any $0 \leq t < k$. Let
\[ z' = y1^k \in \{0,1,2\}^{L(a)}. \]
Then $z'$ is a valid derivation. Take $p'$ such that $p' \equiv z$. Then a similar argument to \eqref{eqarg1} gives $f(a) \leq f(p') < f(c)$ as required. By symmetry $d^a_l > d^b_l \implies f(a) > f(b)$, completing the proof.
\qed
\end{pf}

We have now shown the following. Given two non-negative integers $a$ and $b$ represented with $n$ digits, we can determine the relative order of $a$ and $b$ by scanning the digits in {\em descending\/} order. We can also determine the relative order of $f(a)$ and $f(b)$ by scanning the digits in {\em ascending\/} order. This duality is crucial to the proof that $f$ is a proximity inversion function for the constraint function $\frac{1}{n}$.


\section{Main Theorem}

We now prove that $f$ is a proximity inversion function for the constraint function $\frac{1}{n}$.

\begin{thm}\label{bigspan}
Take $a,b \in \Nset$ such that $a\neq b$ and $f(b) > f(a)$. Then $f(b)-f(a) \geq \frac{1}{\abs{b-a}}$.
\end{thm}

\begin{pf}
Choose any $b \in \Nset$. Now choose a value of $a$ satisfying the requirements of the theorem that maximizes the value of $f(a) + \frac{1}{\abs{b-a}}$. Since $f(b) > f(a) \geq 0$, note that we must have $b > 0$. To prove the theorem, it is sufficient to show that
\begin{equation}\label{mainthmcond}
f(a) + \frac{1}{\abs{b-a}} \leq f(b).
\end{equation}

Let $n = \max\{L(a), L(b)\}$. Since $b > 0$ we must have $n > 0$. We proceed by induction on $n$.
If $n=1$ then $a, b \leq 2$, so the result is easily proved by exhaustion.

For the inductive step, suppose the result is true for $n-1$. Choose $x_a, x_b \in \{0,1,2\}^n$ that satisfy $a \equiv x_a$ and $b \equiv x_b$. We know that $x_a$ and $x_b$ exist, since $n \leq L(a)$ and $n \leq L(b)$. Let $y$ be the longest common prefix of $x_a$ and $x_b$, and let $k = \abs{y} + 1 \leq n$. Since $f(b) > f(a)$ we must have that
\[ d^a_k < d^b_k \]
by Lemma \ref{invmax}. Note that this implies that $L(b) \geq k$. Since $x_a$ is a valid representation and $y$ is a prefix of $x_a$, $y$ must be valid as well. Let $c$ be the non-negative integer satisfying $c \equiv y$.

We shall complete the proof of this theorem in two steps. First, we eliminate all but a few possible representations for $a$ and $b$ by using our results on relative ordering (Property \ref{maxdig} and \thmref{invmax}). We then handle the remaining special cases by using properties of the Fibonacci Sequence.

The first step of the proof depends on the following lemma.
\begin{lem}\label{nobetween}
Suppose $a$ and $b$ are as defined above and there exists $d$ satisfying $f(a) < f(d) < f(b)$. Then either $d > a,b$ or $d < a,b$.
\end{lem}
\begin{pf}
Suppose not; then $d$ is between $a$ and $b$, so 
\[ L(d) \leq \max\{L(a), L(b)\} = n \]
and
\[ \abs{d-a} + \abs{b-d} = \abs{b-a} \]
and hence 
\[ \abs{b-d} < \abs{b-a}. \]
But we then have
\[ f(d) + \frac{1}{\abs{b-d}} > f(a) + \frac{1}{\abs{b-d}}
> f(a) + \frac{1}{\abs{b-a}} \]
which contradicts the maximality of $f(a) + \frac{1}{\abs{b-a}}$.
\qed
\end{pf}

We now use \lemref{nobetween} to eliminate all but a few possible values for $x_a$ and $x_b$. \lemref{nobetween} is a condition on relative ordering, so we can use our results on relative ordering to reduce \lemref{nobetween} to a condition on the characters of $x_a$ and $x_b$.  If, given $x_a$ and $x_b$, we can find a valid $x_d \in \{0,1,2\}^n$ such that (where $r_i \in \{0,1,2\}$ and $w_i, z_i \in \{0,1,2\}^*$)
\begin{enumerate}
\renewcommand{\labelenumi}{(\roman{enumi})}
\item $x_a = w_1r_1z_1$ and $x_d = w_2r_2z_1$ with $r_2 > r_1$; and
\item $x_b = w_3r_3z_2$ and $x_d = w_4r_4z_2$ with $r_3 < r_4$; and
\item $x_a = z_3r_5w_5$, $x_b = z_4r_6w_6$, and $x_d = z_3r_7w_7 = z_4r_8w_8$ with $r_5 < r_7$ and $r_6 > r_8$ (or $r_5 > r_7$ and $r_6 < r_8$),
\end{enumerate}
then taking $d \equiv z$ we arrive at a contradiction via Proposition \ref{maxdig}, \lemref{invmax} and \lemref{nobetween}.

\begin{exmp}
If we had $n = 4$, $x_a = 1112$, and $x_b = 2110$ then we could take $d \equiv 1211$ to arrive at a contradiction.
\end{exmp}

For the remainder of this proof, taking $d \equiv x_d$ will be considered shorthand for this contradiction argument.

\begin{lem}\label{killmost}
Suppose that \eqref{mainthmcond} does not hold. Then $d^b_k = d^a_k + 1$, and $x_a$, $x_b$ must take one of the following forms:

\begin{tabular}[t]{lll}
1. & $x_a = yd^a_k21^*0$ & $x_b = yd^b_k0^*1$\\
2. & $x_a = yd^a_k1^*20$ & $x_b = yd^b_k0^*1$\\
3. & $x_a = yd^a_k21^*01^*20$ & $x_b = yd^b_k0^*1$\\
4. & $x_a = yd^a_k21^*$ & $x_b = yd^b_k0^*$\\
5. & $x_a = yd^a_k0^*1$ & $x_b = yd^b_k1^*20$
\end{tabular}
\end{lem}

\begin{pf}
We proceed by cases based on the values of $L(a)$ and $L(b)$.

\begin{case} 
$L(a) < n-1$, $L(b) = n$. 
\end{case}

Let $t = L(a)$. Then $x_a = yd^a_kw_10^{n-t}$ and $x_b = yd^b_kw_21$ or $x_b = yd^b_kw_22$ for some $w_1, w_2 \in \{0,1,2\}^*$. In either case, take $d \equiv yd^a_kw_10^{n-t-2}10$ to cause a contradiction. We conclude that this case cannot be satisfied.

\begin{case}
$L(a) = n-1$, $L(b) = n$.
\end{case}

That is, $x_b$ ends in 1 or 2 and $x_a$ ends in 10 or 20.

First, suppose $n = k$. Then either $x_b = y1$ or $x_b = y2$. We must also have $x_a = y0$, i.e. $a = c$. If $x_b=y1$ then
\[
f(a) + \frac{1}{\abs{b-a}} = f(c) + \frac{1}{c+u_n-c}
= f(c) + \frac{1}{u_n}
= f(b),
\]
and if $x_b=y2$ then
\[
f(a) + \frac{1}{\abs{b-a}} = f(c) + \frac{1}{c+2u_n-c}
< f(c) + \frac{2}{u_n}
= f(b),
\]
so in either case \eqref{mainthmcond} holds.

Now suppose $n = k+1$. Then $x_a = y10$ and $x_b = y21$. But then take $d \equiv y20$ for contradiction. So we can assume $n > k+1$.

We now show that $d^b_k = d^a_k - 1$.
Well, otherwise $x_a = y0w_1$ and $x_b = y2w_1$. Take $d \equiv y10^{n-k-1}1$ to cause a contradiction.

Consider the string $x_b$. If $x_b = yd^b_k0^trw$ for some $0 \leq t < n - k - 1$, $r \in \{1,2\}$, and $w \in \{0,1,2\}^*$, then take $d \equiv yd^b_k0^{n-k-1}1$. If $x_b = yd^b_k0^{n-k-1}2$ then again take $d \equiv yd^b_k0^{n-k-1}1$.

The only possible value left for $x_b$ is $yd^b_k0^{n-k-1}1$.

Finally, we claim that $x_a$ is of one of the following forms:
\begin{enumerate}
\renewcommand{\labelenumi}{(\roman{enumi})}
\item $yd^a_k21^*0$
\item $yd^a_k1^*20$
\item $yd^a_k21^*01^*20$
\end{enumerate}
To show this, we simply exhaust all other possibilities. The argument is summarized in table \ref{claim13table}. In the strings given, $p,q,r$ refer to arbitrary non-negative integers and $w_i$ refer to arbitrary strings in $\{0,1,2\}^*$.

\begin{table}
\begin{center}
\caption{Exhaustion of all but a few possible representations of $a$ for case 2.}
\begin{tabular}{l|lp{2in}}	\hline
$\mathbf{x_a}$ 	& 	$\mathbf{d}$
\\ \hline \hline
$yd^a_k0w_10$ & $yd^a_k1w_10$
\\ \hline
$yd^a_kw_100$ & $yd^a_kw_110$
\\ \hline
$yd^a_k11^p10$ & $yd^a_k11^p20$
\\ \hline
$yd^a_k11^p0w0$ & $yd^a_k11^p1w0$
\\ \hline
$yd^a_k11^p21^q0w0$ & $yd^a_k21^p11^q1w0$
\\ \hline
$yd^a_k11^p21^q1w0$ & $yd^a_k21^p01^q2w0$
\\ \hline
$yd^a_k21^p01^q0w0$ & $yd^a_k21^p11^q0w0$
\\ \hline
$yd^a_k21^p01^q0w_10$ & $yd^a_k21^p11^q1w_20$
\\ \hline
$yd^a_k21^p01^q21^r10$ & $yd^a_k21^p11^q01^r20$
\\ \hline
$yd^a_k21^p01^q21^r0w0$ & $yd^a_k21^p11^q01^r1w0$
\\ \hline
\end{tabular}
\label{claim13table}
\end{center}
\end{table}

\begin{case}
$L(a) = n$, $L(b) < n-1$.
\end{case}
So $x_a$ ends in $1$ or $2$ and $x_b$ ends with $00$. Note that we must therefore have $n > k+1$, since $d^b_k > 0$.

We begin by showing that $d^b_k = d^a_k + 1$. Otherwise, $d^b_k = 2$ and $d^a_k = 0$. Take $d \equiv y10^{n-k-2}10$ to arrive at a contradiction.

Now consider the string $x_b$. Suppose $x_b \neq yd^b_k0^{n-k}$ for any $k$. Then $x_b = yd^b_k0^trw00$ for some $t \geq 0$, $r \in \{1,2\}$, and $w \in \{0,1,2\}^*$. We take $d \equiv yd^b_k0^{n-k-2}10$ for contradiction.

Finally, we wish to show that $x_a = yd^a_k21^{n-k}$. We simply exhaust all other possibilities. If $x_a = yd^a_k1w$ or $x_a = yd^a_k0w$, take $d \equiv yd^a_k20^{n-k-1}$. If $x_a = yd^a_k21^t0w$, take $d \equiv yd^a_k21^t10^{n-k-t-2}$. So by exhaustion we must have $x_a = yd^a_k21^{n-k}$.

\begin{case}
$L(a) = n, L(b) = n-1$.
\end{case}
Then $x_a$ ends in $1$ or $2$ and $x_b$ ends in $10$ or $20$. Note that we must therefore have $n > k$.

We begin by showing that $d^b_k = d^a_k + 1$. Otherwise $d^b_k = 2$ and $d^a_k = 0$. Take $d \equiv y10^{n-k-1}1$ to arrive at a contradiction.

Now consider the special case $k = n-2$. Then $x_b = yd^b_k0$ and $x_a = yd^a_k1$ or $x_a = yd^a_k1$. In either case, we have
\[ \abs{a-b} \leq 2u_n - u_{n-1} = 2F_{2n} - F_{2n-2} = F_{2n+1} \]
and
\begin{equation*}
\begin{split}
f(b) - f(a) & \geq \frac{1}{u_{n-1}} - \frac{2}{u_n}\\
& = \frac{1}{F_{2n-2}} - \frac{2}{F_{2n}}\\
& = \frac{F_{2n-3}}{F_{2n-1}^2 - 1}\\
& \geq \frac{1}{F_{2n+1}} \quad\text{(by Corollary \ref{catfrac})}\\
& = \frac{1}{\abs{a-b}},
\end{split}
\end{equation*}

so \eqref{mainthmcond} holds. We can now assume that $k < n-2$, so $x_b = yd^b_kwr0$ for some $w \in \{0,1,2\}^*$, $r \in \{0,1,2\}$.

We now claim that $x_a = yd^a_k0^{n-k-1}1$. If not, we must have $x_a = yd^a_k0^{n-k-1}2$ or $x_a = yd^a_k0^trw$ for some $t < n-k-1$ and $r \in \{1,2\}$. Let $d \equiv yd^b_k0^{n-k-1}1$ to arrive at our usual contradiction.

We also claim that $x_b = yd^b_k1^{n-k-2}20$. We simply exhaust all other possibilities. The argument is summarized in Table \ref{claim33table}. In the strings given, $p,q,r$ refer to arbitrary non-negative integers and $w_i$ refer to arbitrary strings in $\{0,1,2\}^*$.
The last case in the table ($x_b = yd^b_k0w$) requires special attention. Note that $yd^a_k$ does not have a suffix of the form $21^*$, since if it did then $yd^b_k$ would have a suffix of the form $21^*2$ contradicting $x_b$'s validity. Thus $yd^a_k1w$ is valid. The fact that $f(a) < f(d)$ follows because $x_a$ has $yd^a_k0$ as a prefix, and $a > d$ follows because $w$ must have suffix $0$ whereas $x_a$ ends with $1$.

\begin{table}
\begin{center}
\caption{Exhaustion of all but a few possible representations of $b$ for case 4.}
\label{claim33table}
\begin{tabular}{l|lp{2in}}	\hline
{\em $x_b$} 	& 	{\em $d$}
\\ \hline
$yd^b_k1^p10w$ & $yd^b_k1^p01w$
\\ \hline
$yd^b_k1^p20w$ & $yd^b_k1^p11w$
\\ \hline
$yd^b_k1^p21w$ & $yd^b_k1^p12w$
\\ \hline
$yd^b_k1^p11$ & $yd^b_k1^p02$
\\ \hline
$yd^b_k0w$ & $yd^a_k1w$
\\ \hline
\end{tabular}
\end{center}
\end{table}

\begin{case}
$L(a) = L(b) = n$.
\end{case}
Then $d^a_n > 0$ and $d^b_n > 0$. Suppose for contradiction that
\begin{equation}\label{contramain}
f(a) + \frac{1}{\abs{b-a}} > f(b).
\end{equation}
Then since we took to $b$ to be arbitrary, we can take the value of $b$ that minimizes $f(b)$, subject to the conditions that there exists some $a$ satisfying \eqref{contramain} and $L(a) = L(b) = n$ (recall that $a$ depends on $b$).
If we consider $a' = a - u_n$, $b' = b - u_n$, we see that 
\[ f(b') = f(b) - \frac{1}{u_n} < f(b) \]
and
\[ f(a') + \frac{1}{\abs{b'-a'}} = f(a) - \frac{1}{u_n} + \frac{1}{\abs{(b-u_n) - (a-u_n)}} > f(b) - \frac{1}{u_n} = f(b'), \]
contradicting the minimality of $f(b)$. 

We conclude that all $a$ and $b$ in this case must satisfy \eqref{mainthmcond}.

This ends the proof of \lemref{killmost}.
\qed
\end{pf}



We must now handle the five cases for $x_a$ and $x_b$ not covered by \lemref{killmost}. We shall handle these remaining cases by appealing to the Fibonacci inequalities developed in Section \ref{sec:fib}.


\begin{lem}
Condition \eqref{mainthmcond} holds if $d^b_k = d^a_k + 1$, $x_b = yd^b_k0^*1$, and $x_a$ takes one of the following forms:
\begin{enumerate}
\item $x_a = yd^a_k21^*0$
\item $x_a = yd^a_k1^*20$
\item $x_a = yd^a_k21^*01^*20$
\end{enumerate}
\end{lem}

\begin{pf}
The cases for $x_a$ can be rewritten as $\,\exists j$, $k < j < n$, such that
\[ a = c + d^a_ku_k + \sum_{i=k+1}^{n-1}u_i + u_{k+1} + u_{n-1} - u_j \]
and hence
\[
f(a) = f(c) + \frac{d^a_k}{u_k} + \sum_{i=k+1}^{n-1}\frac{1}{u_i} + \frac{1}{u_{k+1}} + \frac{1}{u_{n-1}} - \frac{1}{u_j}.
\]
Note that if $n = k+2$ we take $j = k+1$. We also have 
\begin{equation*}
\begin{split}
b & = c + d^b_ku_k + u_n\\
& = c + d^a_ku_k + u_k + u_n.
\end{split}
\end{equation*}

But now Corollary \ref{fibsumsimplecor1} implies that $b-a = u_j$. We therefore have
\begin{equation*}
\begin{split}
& f(a) + \frac{1}{\abs{b-a}}\\
= &\biggl(f(c) + \frac{d^a_k}{u_k} + \sum_{i=k+1}^{n-1}\frac{1}{u_i} + \frac{1}{u_{k+1}} + \frac{1}{u_{n-1}} - \frac{1}{u_j}\biggr) + \frac{1}{u_j}\\
\leq & f(c) + \frac{d^a_k}{u_k} + \frac{1}{u_k} + \frac{1}{u_n} \quad\text{(by Corollary \ref{fibsumcor1})}\\
= & f(c) + \frac{d^b_k}{u_k} + \frac{1}{u_n}\\
= & f(b),
\end{split}
\end{equation*}
as required.
\qed
\end{pf}

\begin{lem}
Condition \eqref{mainthmcond} holds if $d^b_k = d^a_k + 1$, $x_a = yd^a_k21^*$, and $x_b = yd^b_k0^*$.
\end{lem}

\begin{pf}
The conditions on $x_a$ and $x_b$ can be rewritten as
\[ a = c + d^a_ku_k + 2u_{k+1} + \sum_{i=k+2}^nu_i \]
and 
\[ b = c + (d^a_k+1)u_k. \]
So
\[ a-b = 2u_{k+1} + \sum_{i=k+2}^nu_i - u_k = F_{2n+1} \]
by Corollary \ref{fibsumsimplecor2}. But now
\begin{equation*}
\begin{split}
& f(a) + \frac{1}{\abs{a-b}}\\
= & \biggl(f(c) + \frac{d^a_k}{u_k} + \sum_{i=k+1}^n\frac{1}{u_i} + \frac{1}{u_{k+1}}\biggr) + \frac{1}{F_{2n+1}} \\
\leq & f(c) + \frac{d^a_k}{u_k} + \frac{1}{u_k} \quad\text{(by Corollary \ref{fibsumcor2})}\\
= & f(c) + \frac{d^b_k}{u_k}\\
= & f(b),
\end{split}
\end{equation*}
as required.
\qed
\end{pf}

\begin{lem}
Condition \eqref{mainthmcond} holds if $d^b_k = d^a_k + 1$, $x_a = yd^a_k0^*1$, and $x_b = yd^b_k1^*20$.
\end{lem}
\begin{pf}
These conditions on $x_a$ and $x_b$ can be rewritten as
\[ a = c + d^a_ku_k + u_n \]
and 
\begin{equation*}
\begin{split}
b & = c + d^b_ku_k + \sum_{i=k+1}^{n-1}u_k + u_{n-1} \\
& = c + d^a_ku_k + \sum_{i=k}^{n-1}u_k + u_{n-1}.
\end{split}
\end{equation*}
So then
\[ a-b = u_n - u_{n-1} - \sum_{i=k}^{n-1}u_i = F_{2k+1} \]
by Corollary \ref{fibsumsimplecor3}.
We now have 
\begin{equation*}
\begin{split}
f(a) + \frac{1}{\abs{a-b}} & = (f(c) + \frac{d^a_k}{u_k} + \frac{1}{u_n}) + \frac{1}{F_{2k+1}}\\
& \leq f(c) + \frac{d^a_k}{u_k} + \sum_{i=k+1}^{n-1}\frac{1}{u_i} + \frac{1}{u_{n-1}}\\
& < f(c) + \frac{d^b_k}{u_k} + \sum_{i=k+1}^{n-1}\frac{1}{u_i} + \frac{1}{u_{n-1}}\\
& = f(b)
\end{split}
\end{equation*}
by \lemref{fibsum2}, as required.
\qed
\end{pf}

So, in all cases, condition \eqref{mainthmcond} holds. This concludes the proof of \thmref{bigspan}.
\end{pf}

\begin{cor}
$\abs{f(a+n) - f(a)} \leq \frac{1}{a}$ for all $a \geq 0$ and $n \geq 1$.
\end{cor}

\section{Supremum}

We now calculate the supremum of our constructed function $f$.

\begin{thm}
\[
\sup_{i \geq 0}f(i) = 1 + \sum^\infty_{\,n \geq 1}\frac{1}{F_{2n}}.
\]
\end{thm}

\begin{pf}
Choose $n \in \Nset$ and suppose $\displaystyle\max_{L(i) \leq n}f(i)$ is achieved at $a$. Then for all $b \neq a$ with $L(b) \leq n$ we must have $f(b) \leq f(a)$, so by \thmref{invmax} there must exist some $j$, $0 \leq j \leq n$ such that $d^a_j > d^b_j$ and $d^a_i = d^b_i$ for all $0 \leq i < j$. But such a $j$ cannot exist if $b \equiv 21^{n-1}$ (by validity). We must therefore have $a \equiv 21^{n-1}$. Then
\[
\max_{L(i)\leq n}\{f(i)\} = f(a) = \frac{1}{u_1} + \sum_{i=1}^n\frac{1}{u_i} = 1 + \sum_{i=1}^n\frac{1}{F_{2i}}.
\]

Now it is well known that $\frac{F_{2n+2}}{F_{2n}} > \phi^2$ for all $n \geq 1$. Since $F_2 = 1 = \phi^0$, it follows that $F_{2n} > (\phi^2)^{n-1}$ for all $n \geq 1$. We therefore have
\begin{equation*}
\begin{split}
\displaystyle\sum^n_{i=1}\frac{1}{F_{2i}}& < \displaystyle\sum^n_{i=1}\biggl(\frac{1}{\phi^2}\biggr)^{i-1}\\
& < \displaystyle\sum^\infty_{i=0}\biggl(\frac{1}{\phi^2}\biggr)^i\\
& = \frac{\phi^2}{\phi^2-1}\\
& = \phi
\end{split}
\end{equation*}
for all $n \geq 1$. We conclude that $\displaystyle\sum^\infty_{i=1}\frac{1}{F_{2i}}$ converges. So
\[
\sup_{i \geq 1}f(i) = \lim_{n \to \infty}\max_{L(i) \leq n}f(i)
= 1 + \sum^\infty_{n=1}\frac{1}{F_{2n}},
\]
as required.
\qed
\end{pf}

\section{Conclusions and Future Work}

We have constructed a solution to a particular instance of the Generalized Constraint Satisfaction Problem. This solution has a supremum of $1 + \displaystyle\sum^\infty_{n=1}\frac{1}{F_{2n}}$, which we believe is optimal. However, this optimality has not yet been proved. We do, however, put forth the following conjecture which would imply the optimality of our limit.

\begin{conj}
Choose $n > 1$ and let $T = \{1,2,\dotsc,F_{2n}\}$. Then there exists an $f\colon T \to \Rset^{\geq 0}$ satisfying $\abs{f(a)-f(b)} \geq \frac{1}{\abs{a-b}}$ for all $a \neq b$, such that
\[
\max_{a \in T}f(a) = 1 + \biggl(\sum_{i=1}^{n-1}\frac{1}{F_{2i}}\biggr) + \frac{1}{F_{2n+1}}
\]
and this bound is the smallest possible.
\end{conj}

Another avenue of future research is to generalize our approach to other proximity inversion functions. Given any numeration system $S = \{u_i\}$, we can consider a function $f_S \colon \Nset \to \Rset^{\geq 0}$ that maps $\displaystyle\sum_{i=1}^\infty d_iu_i$ to $\displaystyle\sum_{i=1}^\infty \frac{d_i}{u_i}$. We therefore pose the following open problems.  Is it true that for {\em any\/} constraint function $c$ there exists a numeration system $S_c$ such that the function $f_{S_c}$ satisfies $\abs{f(a) - f(a+n)} \geq c(n)$?
If not, what are the necessary and sufficient conditions on $c$ for this to be true? 
Is such an $f_{S_c}$ always an optimal solution?

\begin{ack}
This research was supported in part by Canadian Research Grant NSERC CGSM-301930-2004. I would also like to thank Professor Jeffrey Shallit for helpful discussions regarding numeration systems.
\end{ack}

\end{document}